\documentclass{PoS}
\rightline{\parbox{4cm}{\large\rm 
DESY 06-182\\
Edinburgh 2006/28\\
LTH 726
}}

\usepackage{graphicx}
\newenvironment{Eqnarray}%
          {\arraycolsep 0.14em\begin{eqnarray}}{\end{eqnarray}}

\newcommand{\bc}{\begin{center}}
\newcommand{\ec}{\end{center}}
\newcommand{\eq}{\begin{equation}}
\newcommand{\ee}{\end{equation}}
\newcommand{\ea}{\begin{Eqnarray}}
\newcommand{\eea}{\end{Eqnarray}}

\newcommand{\lsim}{\raisebox{-4pt}{$%
    \,\stackrel{\textstyle <}{\sim}\,$}}
\newcommand{\rgi}{\mbox{\tiny $R\!G\!I$}}                
\newcommand{\msbar}{\mbox{\tiny $\overline{MS}$}}        
\newcommand{\half}{\mbox{\small $\frac{1}{2}$}}          
\newcommand{\bare}{\mbox{\tiny $bare$}}                  

\title{Simulating at Realistic Quark Masses: Light quark masses}
\ShortTitle{Light quark masses $\ldots$}

\author{
        Meinulf G\"ockeler$^{a}$, Roger Horsley$^b$,
        \speaker{Yoshifumi Nakamura}$^c$, Dirk Pleiter$^c$, Paul E.~L. Rakow$^d$,
        Gerrit Schierholz$^{ce}$, Thomas Streuer$^f$, Hinnerk St\"uben$^g$ and
        James M. Zanotti$^b$ \\
        \llap{$^a$} Institut f\"ur Theoretische Physik,
                    Universit\"at Regensburg, \\
                    D-93040 Regensburg, Germany \\
        \llap{$^b$} School of Physics, University of Edinburgh, \\
                    Edinburgh EH9 3JZ, UK \\
        \llap{$^c$} John von Neumann Institute NIC / DESY Zeuthen, \\
                    D-15738 Zeuthen, Germany \\
        \llap{$^d$} Department of Mathematical Sciences,
                    University of Liverpool, \\
                    Liverpool L69 3BX, UK \\
        \llap{$^e$} Deutsches Elektronen-Synchrotron DESY, \\
                    D-22603 Hamburg, Germany \\
        \llap{$^f$} Department of Physics and Astronomy, University of Kentucky,\\
                    Lexington KY 40506, USA \\
        \llap{$^g$} Konrad-Zuse-Zentrum f\"ur Informationstechnik Berlin, \\
                    D-14195 Berlin, Germany \\
        E-mail: \email{meinulf.goeckeler@physik.uni-regensburg.de},
                \email{rhorsley@ph.ed.ac.uk},
                \email{yoshifumi.nakamura@desy.de},
                \email{dirk.pleiter@desy.de},
                \email{rakow@amtp.liv.ac.uk},
                \email{gsch@mail.desy.de},
                \email{thomas.streuer@desy.de},
                \email{stueben@zib.de},
                \email{jzanotti@ph.ed.ac.uk}  }

\author{QCDSF--UKQCD Collaboration}

\abstract{
We present new results for light quark masses.
The calculations are performed using two flavours of $O(a)$ improved Wilson fermions.
We have reached lattice spacings as small as $a \sim 0.07$fm 
and pion masses down to $m_{\pi} \sim 340$MeV in our simulations.
This gives us significantly better control on the
chiral and continuum extrapolations.}

\FullConference{XXIVth International Symposium on Lattice Field Theory\\
                July 23-28, 2006\\
                Tucson, Arizona, USA}
\begin{document}


\section{Introduction}
The `running' of the renormalised quark mass as the scale $M$
is changed is controlled by the $\beta$ and $\gamma$ functions
in the renormalisation group equation, defined by
\begin{eqnarray}
   \beta^{ S} \left(g^{ S}(M) \right) &\equiv&
                 \left. {\partial g^{ S}(M) \over
                         \partial \log M }\right|_{\bare},
                                   \label{beta_def} \\
   \gamma_m^{ S} \left(g^{ S}(M) \right) &\equiv&
                 \left. {\partial \log Z_m^{ S}(M) \over
                         \partial \log M }\right|_{\bare},
                                   \label{gamma_def}
\end{eqnarray}
where the bare parameters are held constant. These functions are given
perturbatively as power series expansions in the coupling constant,
\begin{eqnarray}
   \beta^{ S}(g)    &=& - b_0g^3 - b_1g^5
                            - b_2^{ S}g^7 - b_3^{ S}g^9 - \ldots \,,
                                    \nonumber \\
   \gamma_m^{ S}(g) &=&   d_{m0}g^2 + d_{m1}^{ S}g^4
                            + d_{m2}^{ S}g^6 + d_{m3}^{ S}g^8 +
                                          \ldots \,.
\end{eqnarray}
The first two coefficients of the $\beta$-function and first coefficient
of the $\gamma_m$ function are scheme independent,
\begin{equation}
   b_0 = {1\over (4\pi)^2}
           \left( 11 - {2\over 3}n_f \right) \,, \qquad
   b_1 = {1\over (4\pi)^4}
           \left( 102 - {38 \over 3} n_f \right) \,.
\label{b0+b1}
\end{equation}
and
\begin{equation} 
   d_{m0} = - { 8 \over (4\pi)^2} \,,
\end{equation}
while all others depend on the scheme chosen.

We may immediately integrate eq.~(\ref{beta_def}) to obtain
\begin{equation}
   { M \over\Lambda^{ S} }
      =  \left[b_0 g^{ S}(M)^2
                          \right]^{b_1\over 2b_0^2}
         \exp{\left[{1\over 2b_0 g^{ S}(M)^2}\right]} 
         \exp{\left\{ \int_0^{g^{ S}(M)} d\xi
          \left[ {1 \over \beta^{ S}(\xi)} +
                 {1\over b_0 \xi^3} - {b_1\over b_0^2\xi}
         \right]\right\} } \,.
\label{lambda_def}
\end{equation}
The renormalisation group invariant (RGI) quark mass%
\footnote{Analogous definitions hold for other quantities which depend
on the scheme and scale chosen.}
is defined from the renormalised quark mass as
\begin{equation}
   m_q^{\rgi} \equiv \Delta Z_m^{ S}(M) m^{ S}(M)
               = \Delta Z_m^{ S}(M) Z_m^{ S}(M) m_q^{\bare}
               \equiv Z_m^{\rgi} m_q^{\bare} \,,
\label{mrgi_msbar}
\end{equation}
where
\begin{equation}
   [\Delta Z_m^{ S}(M)]^{-1} = 
          \left[ 2b_0 g^{ S}(M)^2 \right]^{-{d_{m0}\over 2b_0}}
          \exp{\left\{ \int_0^{g^{ S}(M)} d\xi
          \left[ {\gamma_m^{ S}(\xi)
                             \over \beta^{ S}(\xi)} +
                 {d_{m0}\over b_0 \xi} \right] \right\} },
\label{deltam_def}
\end{equation}
and so the integration constant upon integrating eq.~(\ref{beta_def})
is given by $\Lambda^{ S}$, and similarly from eq.~(\ref{gamma_def})
the integration constant is $m_q^{\rgi}$. $\Lambda^{ S}$ and
$m_q^{\rgi}$ are thus independent of the scale. (Note that although
the functional form of $\Delta Z_m^{ S}(M)$ is fixed,
the absolute value is not; conventions vary for its definition.)
Also for a scheme change ${ S}\to { S}^\prime$
(it is now sufficient to take them at the same scale) given by
\begin{equation}
   g^{{ S}^\prime} = G(g^{ S}) = 
             g^{ S}(1 + \half t_1 (g^{ S})^2 + \ldots) \,,
\label{G_def}
\end{equation}
$m_q^{\rgi}$ remains invariant, while $\Lambda$ changes as
$\Lambda^{{ S}^\prime} = \Lambda^{ S} \exp(t_1/(2b_0))$.
Note also that analytic expressions for the integrals in 
eq.~(\ref{mrgi_msbar}) or eq.~(\ref{deltam_def}) can be found for
low orders, for example to two loops we have
\begin{equation}
   \Delta Z^{ S}_m(M) =
      \left[ 2 b_0 (g^{ S}(M))^2 \right]^{d_{m0}\over 2b_0}
      \left[ 1 + {b_1 \over b_0} (g^{ S}(M))^2
      \right]^{{b_0 d_{m1}^{ S} - b_1 d_{m0}
                                               \over 2 b_0 b_1} } \,.
\label{DelZ_2loop}
\end{equation}
Thus we have a convenient splitting of the problem into two parts:
a number, $m_q^{\rgi}$, which involves a non-perturbative computation,
and is the goal of this paper and, if desired, an evaluation of
$\Delta Z_m^{ S}$ which allows the running quark mass to be given
in a renormalisation scheme ${ S}$.
\section{Simulation}
We have estimated the light quark masses in the $\overline{MS}$ scheme at 2GeV
    by first using the axial Ward identity (AWI) to determine the
    lattice quark mass. This is renormalised using the $\rm{RI}^\prime - \rm{MOM}$
    scheme~\cite{martinelli94a} (for our variation on the method see~\cite{hep-lat/9807044}),
    converted to a RGI form as described in section~1 and after the continuum
    limit has been taken rewritten in the $\overline{MS}$ scheme.
    Further details and results are given in~\cite{MG:2006qmass}.
We perform our simulations with two flavours of non-perturbatively
clover-im\-proved dynamical Wilson fermions and Wilson glue. Using these
actions, the QCDSF and UKQCD collaborations have generated gauge field
configurations with the parameters given in Table~\ref{tab:lats}.
We also use configurations generated by the DIK collaboration which have been
made available through the ILDG.
This large set of lattices enables us to extrapolate to the chiral and the continuum limit.
\begin{table}[!bt]
  \caption{\label{tab:lats} Overview of our lattice parameters. For the translation into physical
    units the Sommer parameter\cite{sommer93a} with $r_0=0.467$~fm 
(see \cite{Khan:2006de} and \cite{Aubin:2004wf})
 has been used.
}
  \begin{center}
  \begin{tabular}{ccccccc}
    \hline\hline%
$\beta$ & $\kappa$& $N^3 \times  T$ &$m_{\pi}$~[GeV]  & $a$~[fm] & $L$~[fm]& $N_{traj}$\\ \hline
   5.20 & 0.13420 & $16^3\times 32$ &1.007(2) & 0.115 & 1.8 &       $\mathcal{O}(5000)$\\
        & 0.13500 & $16^3\times 32$ &0.833(3) & 0.098 & 1.6 &       $\mathcal{O}(8000)$\\
        & 0.13550 & $16^3\times 32$ &0.619(3) & 0.093 & 1.5 &       $\mathcal{O}(8000)$\\ \hline
   5.25 & 0.13460 & $16^3\times 32$ &0.987(2) & 0.099 & 1.6 &       $\mathcal{O}(6000)$\\
        & 0.13520 & $16^3\times 32$ &0.829(3) & 0.091 & 1.5 &       $\mathcal{O}(8000)$\\
        & 0.13575 & $24^3\times 48$ &0.597(1) & 0.084 & 2.0 &       $\mathcal{O}(6000)$\\ \hline
   5.29 & 0.13400 & $16^3\times 32$ &1.173(2) & 0.097 & 1.6 &       $\mathcal{O}(4000)$\\
        & 0.13500 & $16^3\times 32$ &0.929(2) & 0.089 & 1.4 &       $\mathcal{O}(5600)$\\
        & 0.13550 & $24^3\times 48$ &0.769(2) & 0.084 & 2.0 &       $\mathcal{O}(2000)$\\
        & 0.13590 & $24^3\times 48$ &0.591(2) & 0.080 & 1.9 &       $\mathcal{O}(5000)$\\
        & 0.13620 & $24^3\times 48$ &0.395(3) & 0.077 & 1.9 &       $\mathcal{O}(3000)$\\
        & 0.13632 & $32^3\times 64$ &0.337(3) & 0.077 & 2.5 &       $\mathcal{O}(1000)$\\ \hline
   5.40 & 0.13500 & $24^3\times 48$ &1.037(1) & 0.077 & 1.8 &       $\mathcal{O}(4000)$\\
        & 0.13560 & $24^3\times 48$ &0.842(2) & 0.073 & 1.8 &       $\mathcal{O}(3000)$\\
        & 0.13610 & $24^3\times 48$ &0.626(2) & 0.070 & 1.7 &       $\mathcal{O}(4000)$\\
        & 0.13640 & $24^3\times 48$ &0.432(3) & 0.068 & 1.6 &       $\mathcal{O}(3000)$\\
    \hline\hline
  \end{tabular}
  \end{center}
\label{tab:lats}
\end{table}
\section{Quark masses}

\begin{figure}[!thb]
\bc
\vspace{-4mm}
\includegraphics[angle=270,scale=0.3,clip=true]{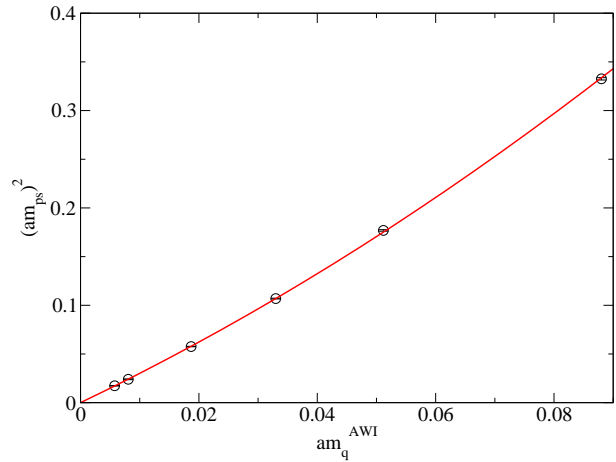}
\ec
\vspace{-4mm}
\caption{
$(am_{ps})^2$ versus $am^{AWI}_{q}$ for $\beta$=5.29.
}
\label{chpt5.29}
\end{figure}

As a first check we perform the chiral extrapolation for pseudoscalar mass. 
In Fig.~\ref{chpt5.29} we plot $(am_{ps})^2$ against $am^{AWI}_{q}$ together
with the fit result.
Our data shows that $(am_{ps})^2$ goes to $0$ at $am^{AWI}_{q}$=0.
This means that even at our lightest quark mass the data is not
 seriously effected by either an Aoki phase or weak 1st order phase
 transition.

We use the next to leading order (NLO) chiral perturbation theory ($\chi$PT)
to estimate the quark masses,
\begin{eqnarray}
   r_0m_s^{\rgi}
       &=& c^{\rgi}_a 
               \left[ (r_0m_{K^+})^2 + (r_0m_{K^0})^2 - (r_0m_{\pi^+})^2 
               \right]
                                           \nonumber \\
       & & + (c^{\rgi}_b-c^{\rgi}_d)
                \left[(r_0m_{K^+})^2 + (r_0m_{K^0})^2 \right](r_0m_{\pi^+})^2 
                                           \nonumber \\
       & & + \half (c^{\rgi}_c+c^{\rgi}_d)
                \left[(r_0m_{K^+})^2 + (r_0m_{K^0})^2 \right]^2
                                           \nonumber \\
       & & - (c^{\rgi}_b+c^{\rgi}_c)(r_0m_{\pi^+})^4
                                           \nonumber \\
       & & - c^{\rgi}_d \left[(r_0m_{K^+})^2 + (r_0m_{K^0})^2 \right]
                 \left[ (r_0m_{K^+})^2 + (r_0m_{K^0})^2 - (r_0m_{\pi^+})^2
                 \right]
                                           \nonumber \\
       & & \hspace*{1.0in} \times
             \ln \left( (r_0m_{K^+})^2 + (r_0m_{K^0})^2 - (r_0m_{\pi^+})^2
                 \right)
                                           \nonumber \\
       & & + c^{\rgi}_d (r_0m_{\pi^+})^4 \ln (r_0m_{\pi^+})^2 \,
\label{strange_r0}
\end{eqnarray}
\begin{equation}
   r_0m^{\rgi}_{ud} = c_a^{\rgi} (r_0m_{\pi^+})^2 
                      + (c_b^{\rgi}+c_c^{\rgi}) (r_0m_{\pi^+})^4
                      - c_d^{\rgi} (r_0m_{\pi^+})^4 \ln (r_0m_{\pi^+})^2 \,.
\label{ud_r0}
\end{equation}
The fit function to determine $c_a^{\rgi}$ and $c_i^{\rgi}$, $i = b, c, d$ is
\begin{equation}
   { r_0 m_q^{\rgi} \over (r_0 m_{ps})^2 }
      = c^{\rgi}_a +
        c^{\rgi}_b (r_0 m^S_{ps})^2 +
        c^{\rgi}_c (r_0m_{ps})^2 +
        c^{\rgi}_d \left( (r_0 m^S_{ps})^2 - 2(r_0 m_{ps})^2 
                               \right) \ln (r_0 m_{ps})^2 \,,
\label{fit_function}
\end{equation}
where $m_{ps}$, $m_{ps}^S$ are the valence and sea pseudoscalar masses respectively
(both using mass degenerate quarks, since we found the relevant quantities
$a m_{ps}$ and $a {m}_q$ to differ by $\lsim 1\%$
between the degenerate quarks case and the non-degenerate quarks case).
The first term is the leading order, LO, result in $\chi$PT while the
remaining terms come from the next non-leading order, NLO, in $\chi$PT.
We note that to NLO, we can determine
$c_a^{\rgi}$ and $c_i^{\rgi}$, $i = b, c, d$ using mass degenerate
quarks and then simply substitute them in eqs.~(\ref{strange_r0}, \ref{ud_r0}).

To reduce the total error on the result, it proved advantageous
to use eq.~(\ref{strange_r0}) to eliminate $c_a^{\rgi}$ from
eq.~(\ref{fit_function}) in terms of
\begin{equation} 
   c^{\rgi}_{a^{\prime}} \equiv
      { r_0m_s^{\rgi} \over
           (r_0m_{K^+})^2 + (r_0m_{K^0})^2 - (r_0m_{\pi^+})^2 } \,.
\end{equation}
This results in a modified fit function of the form
\begin{eqnarray}
   { r_0 m_q^{\rgi} \over (r_0 m_{ps})^2 }
      = c^{\rgi}_{a^{\prime}} &+& 
        c^{\rgi}_b [ (r_0 m^S_{ps})^2 - d_b ] +
        c^{\rgi}_c [ (r_0m_{ps})^2 - d_c ]
                                                     \nonumber \\
            &+& 
        c^{\rgi}_d \left[ \left( (r_0 m^S_{ps})^2 - 2(r_0 m_{ps})^2 
                               \right) \ln (r_0 m_{ps})^2 - d_d \right]\,,
\label{modified_fit_function}
\end{eqnarray}
where $d_i$ ($i=b, c, d$) can be read-off from eq.~(\ref{strange_r0})
and have the effect of shifting the various terms in the fit function
by a constant.

\begin{figure}[!thb]
\bc
\vspace{-3mm}
\includegraphics[angle=270,scale=0.3,clip=true]{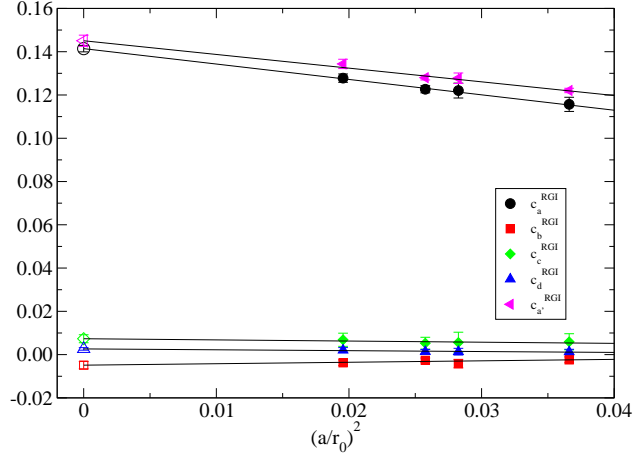}
\ec
\vspace{-4mm}
\caption{
$c_a^{\rgi}$, $c_i^{\rgi}$ ($i = b, c, d$) and $c^{\rgi}_{a^{\prime}}$ versus $(a/r_0)^2$.
            The open symbols represent the
            values of $c_a^{\rgi}$, $c_i^{\rgi}$ ($i = b, c, d$) and $c^{\rgi}_{a^{\prime}}$
            in the continuum limit.
}
\label{c_r0}
\end{figure}

In Fig.~\ref{c_r0} we plot $c_a^{\rgi}$, $c_i^{\rgi}$ ($i = b, c, d$) and $c^{\rgi}_{a^{\prime}}$
 against $(a/r_0)^2$.
The coefficients of NLO are small compared to the LO coefficient. Finally, we find
\begin{equation}
   m_s^{\msbar}(2\,\mbox{GeV})
           = \left\{ \begin{array}{lll}
                        121(2)(3)(6)\,\mbox{MeV} & \mbox{for} &
                          r_0 = 0.5 \,\mbox{fm}   \\
                        115(2)(3)(6)\,\mbox{MeV} & \mbox{for} &
                          r_0 = 0.467 \, \mbox{fm}
                       \end{array}
               \right. \nonumber
\end{equation}
where the first error is statistical and the second is systematic
$\approx 3\,\mbox{MeV}$. We have determined it from the effect on $c_{i}^{\rgi}$ by changing
the fit interval $(r_0 m_{ps})^2 \lsim 5$ to $(r_0 m_{ps})^2 \lsim 4$ or $6$ or $\infty$,
i.e.\ include all the data.
Furthermore the additional third (systematic) error is
due to the uncertainty with which value to identify $r_0$.

For the light quark mass, we find that 
corrections from LO to NLO $\chi$PT are negligibly small.
We shall just quote the LO result of
\begin{equation}
   m_{ud}^{\msbar}(2\,\mbox{GeV})
           = \left\{ \begin{array}{lll}
                        4.57(05)(07)(23)\,\mbox{MeV} & \mbox{for} &
                          r_0 = 0.5 \, \mbox{fm}  \\
                        4.34(05)(07)(23)\,\mbox{MeV} & \mbox{for} &
                          r_0 = 0.467 \, \mbox{fm}
                       \end{array}
               \right. \nonumber
\end{equation}
where again the second and third errors are systematic.
Finally, we see that the ratio
\begin{eqnarray}
  {m_s^{\msbar}(2\,\mbox{GeV}) \over m_{ud}^{\msbar}(2\,\mbox{GeV})}
             = 26.6(1.8).
\end{eqnarray}
\section{Conclusion}
We have updated our estimate for the light quark masses using results 
at smaller lattice spacing and smaller quark masses data.
In Table \ref{results}, we compare the updated and previously published values~\cite{MG:2006qmass}.
Our results are in rough agreement with other group's results.
In order to improve the precision and accuracy of analysis, 
simulations of smaller quark masses and lattice spacing
and 2+1 flavours are needed.

\begin{table}[!bt]
  \caption{\label{results} 
Comparison of the updated and previous
values of 
$m^{\msbar}_s(2 GeV)$, $m^{\msbar}_{ud}(2 GeV)$ and
$m^{\msbar}_s(2 GeV)/m^{\msbar}_{ud}(2 GeV)$.}
      \begin{center}
\vspace{-5mm}
      \begin{tabular}{lll}
\multicolumn{1}{c}{}     &
\multicolumn{1}{c}{previous}&
\multicolumn{1}{c}{new}  \\ 
         \hline
$m^{\msbar}_s(2 GeV)$                        & $111(6)(4)(6)    $MeV & $115(2)(3)(6)    $MeV \\
$m^{\msbar}_{ud}(2 GeV)$                     & $4.08(23)(19)(23)$MeV & $4.34(05)(07)(23)$MeV \\
$m^{\msbar}_s(2 GeV)/m^{\msbar}_{ud}(2 GeV)$ & $27.2(3.2)       $    & $26.6(1.8)       $    \\
         \hline
      \end{tabular}\\
      \end{center}
\label{results}
\end{table}

\acknowledgments
The numerical calculations have been performed on the Hitachi SR8000 at LRZ (Munich),
on the Cray T3E at EPCC (Edinburgh),
on the Cray T3E at NIC (J\"ulich) and ZIB (Berlin),
as well as on the APEmille and apeNEXT at DESY (Zeuthen),
while simulations at the smallest three pion masses have been performed on
the BlueGeneL at NIC/J\"ulich, EPCC at Edinburgh and 
KEK at Tsukuba by the Kanazawa group as part of the DIK research programme.
We thank all institutions.
This work has been supported in part by
the EU Integrated Infrastructure Initiative Hadron Physics (I3HP) under
contract RII3-CT-2004-506078
by the DFG under contract FOR 465 
(Forschergruppe 

\hspace{-8mm}
Gitter-Hadronen-Ph\"anomenologie).
We would also like to thank A.~Irving for providing updated results 
for $r_0/a$ prior to publication.


\begin{thebibliography}{99}

\bibitem{martinelli94a}
G. Martinelli {\em et~al.},
\newblock Nucl. Phys. {\bf B445} (1995) 81, [hep-lat/9411010].

\bibitem{hep-lat/9807044}
M.~G\"ockeler {\em et~al.},
\newblock Nucl. Phys. {\bf B544} (1999) 699, [hep-lat/9807044].

\bibitem{MG:2006qmass}
M.~G\"ockeler {\em et~al.},
\newblock Phys. Rev. {\bf D73}, 054508 (2006), [hep-lat/0601004].

\bibitem{sommer93a}
R. Sommer,
\newblock Nucl. Phys. {\bf B411} (1994) 839, [hep-lat/9310022].

\bibitem{Khan:2006de}
A.~Ali~Khan {\em et~al.},
\newblock hep-lat/0603028.

\bibitem{Aubin:2004wf}
C.~Aubin {\em et~al.},
\newblock Phys. Rev. {\bf D70}, 094505 (2004), [hep-lat/0402030].


\end{thebibliography}
\end{document}